\documentstyle[preprint,aps,amsfonts]{revtex}
\begin{document}

\draft

\title{Mesoscopic Systems With Fixed Number of Electrons}
\author{W. Lehle}
\address{Institut f\"ur Theorie der Kondensierten Materie,
Universit\"at Karlsruhe,\\ 76128 Karlsruhe, Germany}
\author{ A. Schmid\cite{AS}}
\address{Department of Condensed Matter Physics, Weizmann
Institute of Science,\\ Rehovot 76100, Israel}
\date{\today}

\maketitle

\begin{abstract}
In this paper, we study the physics of mesoscopic systems with
noninteracting, but  fixed number of electrons.   From a technical
point of view, this means a discussion of the differences between the
canonical and the grand canonical ensemble (fixed versus fluctuating
number of particles).  Such a discussion is not trivial since the
grand canonical ensemble is the most convenient basis for the
statistics of identical particles and one has to spend labour
in order to retrieve the canonical ensemble. Specifically, we are
considering
ensembles of mesoscopic systems with disorder,
either by atomic defects or by fluctuations in their geometric
definitions and we discuss various forms of disorder averages.
\end{abstract}

\section{INTRODUCTION}
\label{sec:intro}

In the following, we will investigate some  properties of
noninteracting  electrons in mesoscopic systems.
We take into account that irregularities in the
preparation introduce  disorder on an atomic scale and also, in the
geometrie definition of the samples.  We will explore
the consequences of strict conservation of particle number and we will
find that some details in the disorder averaging procedure may become
important.  Such a peculiar behavior has been emphasized by
various authors; we wish to mention here Shklovskii \cite{Shk} as well as
Imry \cite{Imry}. From a technical point of view, our paper is meant
to discuss the {\it differences between the canonical and the grand
canonical statistical ensemble.}

We will be concerned first with (A) thermodynamic properties on the
average \cite{Imry},
secondly with (B) kinetic behavior \cite{Shk}, and thirdly with (C)
stochastic properties of thermodynamic quantities [3].
Clearly, in the first and third case,
 it is suffice to know the positions of the single particle
levels $\epsilon_\lambda$.  We will argue that the position of the
levels is also most decisive  in the second case.

(A) According to standard arguments, thermodynamic properties may be
derived from the grand canonical potential $\Omega (T,\mu)$ which is
given by
\begin{equation}
\Omega = \sum_\lambda (-T)\ell n \Big[1 + e^{-(\epsilon_\lambda -
\mu)/T}\Big]
\label{1}
\end{equation}
We introduce the single particle density of states per spin
\begin{equation}
{\cal D}(E) = {1\over s}\sum_\lambda \delta(E-\epsilon_\lambda)
\label{2}
\end{equation}
where $s = (2S + 1)$ is the spin degeneracy.  (In previous
publications, we used to start with
the density of states per spin and unit volume ${\cal N}(E) = {\cal
D}(E)/{\cal V}$, following a convention of previous
decades, where
one has been interested  in bulk properties.)
Then, the expression for
the grand canonical potential can be written as
\begin{equation}
\Omega = s\int dE {\cal D}(E) (-T)\ell n \Big[1 + e^{-(E-\mu)/T}\Big]
\label{3}
\end{equation}

(B) Following Ref. \cite{Shk}, we will also study the photoabsorption
of mesoscopic particles.  The cross-section for this process can be
written \cite{Shk} in the form
\begin{equation}
\sigma={4\pi^2e^2\over
3c}\sum_{\lambda\lambda{'}}(\epsilon_{\lambda{'}}-\epsilon_\lambda)\mid
R_{\lambda\lambda{'}}\mid^2(n_\lambda - n_{\lambda{'}})\delta
(\hbar\omega - \epsilon_{\lambda{'}}+\epsilon_\lambda)
\label{4}
\end{equation}
where $n_\lambda$ is the population of the single particle state
$\mid\lambda >$; and $eR_{\lambda\lambda{'}} = e <\lambda\mid\hat R\mid
\lambda{'}>$ the matrix element of the electronic dipole moment.  It seems
(see Refs. [4] - [6]), that there is not much structure in $\mid
R_{\lambda\lambda{'}}\mid^2$ so that one may replace it by an average
value
\begin{equation}
\sigma_0 = {4\pi^2e^2\over 3c\hbar}<\mid R_{\lambda\lambda{'}}\mid^2>
\label{5}
\end{equation}
(which  also includes an average with respect to the disorder), without
distorting the general structure of the result to be obtained.
Assuming that the population is equal to the Fermi function, we obtain
\begin{equation}
\sigma=\sigma_0\hbar^2\omega\;\; s\int dEdE{'}{\cal D}(E){\cal D}(E{'})
\Big[f(E-\mu) - f(E{'}-\mu)\Big]\delta (\hbar \omega - E{'}+E)
\label{6}
\end{equation}

Due to the discreteness of the levels in  mesoscopic samples, the
$\omega$-dependence of the cross section $\sigma(\omega)$ consists of
$\delta$-spikes for each individual sample and it is only the average
with respect to the ensemble of samples, that is, the disorder averaged
cross section $<\sigma(\omega)>$, which  may be expected to be a continuous
function of the frequency.

As a rule, the density of states is a fluctuating quantity which
consists of a leading part ${\cal D}^0(E)$ and a remainder
\begin{mathletters}
\label{eq:7}
\begin{equation}
{\cal D}(E) = {\cal D}^0(E) + {\cal D}^1(E)
\label{7a}
\end{equation}
where
\begin{equation}
{\cal D}^0(E) = <{\cal D}(E) >
\label{7b}
\end{equation}
is equal to the average with respect to the disorder. Note that in the
literature, the
following
abbreviation occurs quite frequently
\begin{equation}
{\cal R}_1(E) = {\cal D}^0(E)
\label{7c}
\end{equation}
\end{mathletters}
(one level correlation functions); and if systematics is required,
we will resort to that terminology.

We mention in passing that in some problems where no disorder appears
explicitly, the average can be taken with respect to an energy window
or it
may represent a kind of integration with respect to quantum numbers.

As a rule, ${\cal D}^0$ is insensitive to
genuine quantum phenomena which may  result e.g. from  an applied
magnetic
field or from a magnetic flux threading a mesoscopic ring.

If this is the case, the disorder averaged grand canonical potential
\begin{equation}
\Omega^0 =: <\Omega > = s\int dE {\cal D}^0(E)(-T)\ell n
\Big[1+e^{-(E-\mu)/T}\Big]
\label{8}
\end{equation}
has lost any sensitive dependence on external parameters.  Indeed,
this turns out to be true as far as the response (persistent currents)
to magnetic flux of mesoscopic rings is concerned.  On the other hand,
it is found that there is a measurable effect in a theory where the
number of electrons is kept fixed [2,7].

(C) The fact that the sensitive part of $\Omega$ is, on the average,
zero does not mean that $\Omega$ is zero for each realization.
Rather, it means that its contributions fluctuate in magnitude and
in sign; loosely
speaking, one may say that plus and minus signs occur in about equal
numbers. The simplest quantity which can be used as a measure of the
stochastic (sample to sample) fluctuations, is the mean square value.
In previous publications (see references in [3]), it has been
considered to be sufficient for non-interacting electrons, to
calculate the mean square  $<\Omega^2>$, which means stochastic
fluctuations at fluctuating numbers of electrons.  We will show how
stochastic fluctuations can be calculated for fixed number of
electrons. A preliminary estimate of the resulting expression
indicates that significantly larger values are then obtained as compared
with the standard procedure.

For an initiation to a discussion of the photoabsorption problem, we
calculate  the disorder averaged
cross section in a crude approximation $<\sigma >^{uc}$ where in Eq. (\ref{6})
the averaged product of the density of states is replaced by the
product of the averages ($uc$: uncorrelated; that is, disregarding
correlations).
If $\hbar\omega << \mu$, we may even put
${\cal D}^0(E) \approx {\cal D}^0(\mu^0)$ where
\begin{equation}
{\cal D}^0(\mu^0) =: {\cal D}_F
\label{9}
\end{equation}
is the average density of states at a (typical) Fermi level $\mu^0$.
Then, we obtain
\begin{equation}
<\sigma >^{uc} = \hbar^3\omega^2 s {\cal D}_F^2\sigma_0
\label{10}
\end{equation}

The paper is organized as follows.  Thermodynamic properties (A) of
mesoscopic systems with fixed number of electrons are investigated in
Section II, III, and IV on the basis of (i) Legendre Transformation;
(ii) Coulomb blockade; and (iii) pinning of the Fermi level to a
single particle state.  Section V is devoted to the
problem of photoabsorption, whereas, stochastic fluctuations are
studied in Section VI.  The paper concludes with a discussion in
Section VII. Appendix A presents a calculation of correlation
functions in the cooperon-diffuson approximation for persistent
currents and Appendix B contains a collection of results of the
random matrix theory.

\section{Canonical Ensemble by Legendre Transformation}
\label{sec:can}

The mathematical conveniences provided by the grand canonical ensemble
(fluctuating number of particles) are evident and there is no
need to consult text books.  Certainly, there are
differences in the results obtained for the above mentioned ensemble
and the canonical ensemble (fixed number of particles) but they
are said to become relatively small in the thermodynamic limit.

However, there are phenomena in mesoscopic systems  which
depend only on a relatively small number of electrons.  This suggests
that one should work entirely within the canonical ensemble; in fact,
this
retriction should be taken very serious since in all processes the
particle number is conserved.

At low temperatures (at $T=0$, to be precise) where thermal
fluctuations are negligeable, the grand canonical potential
$\Omega(T = 0,\mu)$ and the free energy $F(T=0,N)$ are directly
related by a Legendre transformation.  However,   this is
true only before the averages with respect to the disorder have been
taken; therefore, the transformation has to be done for each
realization of the disorder separately.  This program can be carried
through comparably easy, on the basis
of the decomposition (7), provided that the fluctuating part ${\cal
D}^1(E)$ in the density of states is small.  We will find that the
first nontrivial correction is of second order
in ${\cal D}^1$. In what follows, we will relax the condition
$T=0$ and we will perform the Legendre transformation from
$\Omega(T,\mu)$ to $F(T,N)$ on the basis of the above mentioned
expansion [7].

In this context, it is convenient to formulate the Legendre
transformation as follows
\begin{equation}
F(T,N) = {\rm max}_\mu \Big\{\Omega(T,\mu) + \mu N\Big\}
\label{11}
\end{equation}
Clearly, a necessary condition  for a maximum is
\begin{equation}
{\partial\Omega\over\partial\mu} + N = 0 \Rightarrow \hskip 1cm
\mu = \mu (T,N)
\label{12}
\end{equation}
Corresponding to the decomposition (7) of the density of states we have
$\Omega = \Omega^0 + \Omega^1$, where $\Omega^0 = <\Omega >$ is given
in Eq. (\ref{8}) and where
\begin{equation}
\Omega^1 = s \int dE {\cal D}^1(E) (-T)\ell n \Big[1 + e^{-(E-\mu)/T}\Big]
\label{13}
\end{equation}
Let us put
\begin{equation}
\mu = \mu^0 + \mu^1
\label{14}
\end{equation}
where $\mu^0$ is the zero order approximation such that
\begin{mathletters}
\label{eq:15}
\begin{equation}
- {\partial\Omega^0\over \partial
\mu}\Biggr|_{_{\displaystyle{\mu_0}}}
= N
\label{15a}
\end{equation}
and let us assume that we have found the leading order Legendre
transform
\begin{equation}
F^0(T,N) = \Omega^0(\mu^0) + \mu^0N
\label{15b}
\end{equation}
\end{mathletters}
According to what we have said in the Introduction, $F^0$ does
not depend on the quantum effects we are interested in; therefore,
its detailed form is not important here.  In the next order, we have
\begin{mathletters}
\label{eq:16}
\begin{equation}
- {\partial\Omega^0\over \partial \mu}\Biggr|_{_{\displaystyle{\mu^0 +
\mu^1}}} - {\partial\Omega^1\over
\partial\mu}\Biggr|_{_{\displaystyle{\mu^0}}} = N
\label{16a}
\end{equation}
Making use of the relation (cf. Eq. (\ref{9}))
\begin{equation}
 - {\partial^2\Omega^0\over
\partial\mu^2}\Biggr|_{_{\displaystyle{\mu^0}}} = {\partial N\over
\partial\mu}\Biggr|_{_{\displaystyle{\mu^0}}} = s{\cal D}_F
\label{16b}
\end{equation}
\end{mathletters}
we obtain
\begin{eqnarray}
s{\cal D}_F && \mu^1 + s\int dE {\cal D}^1(E)f(E-\mu^0) = 0\nonumber
\\
\mu^1 && = {1\over {\cal D}_F}\int dE{\cal D}^1(E)f(E-\mu^0)
\label{18a}
\end{eqnarray}

We conclude that through second order
\begin{equation}
F = F^0 + \Omega^1(\mu^0) + \Delta F
\label{19}
\end{equation}
where
\begin{equation}
\Delta F = {\rm max}_{\mu^1} \Big\{{1\over 2}{\partial^2\Omega^2\over
\partial\mu^2}(\mu^1)^2 + {\partial\Omega^1\over \partial\mu}\mu^1\Big\}
\label{20}
\end{equation}
and where the derivatives are taken at $\mu = \mu^0$.  Consequently,
\begin{eqnarray}
\Delta F && = - {1\over 2}{\partial^2\Omega^0\over
\partial\mu^2}(\mu^1)^2\nonumber\\
&& = {1\over 2} {1\over s{\cal D}_F}\Big[s\int dE {\cal
D}^1(E)f(E-\mu^0)\Big]^2
\label{21}
\end{eqnarray}
For convenience, we introduce here
\begin{equation}
\Delta = 1/s{\cal D}_F
\label{22}
\end{equation}
which is the mean level separation and
\begin{equation}
\delta N = s\int dE {\cal D}^1(E)f(E-\mu^0)
\label{23}
\end{equation}
which is the disorder induced fluctuation in the particle number
calculated for the grand canonical ensemble.  Thus, we may write
\begin{equation}
\Delta F = {1\over 2}\Delta (\delta N)^2
\label{24}
\end{equation}

We recognize that the calculation of the disorder average $<\Delta F>$
requires the knowledge of the  two-level correlation function
\begin{mathletters}
\label{eq:24}
\begin{equation}
\tilde{\cal R}_2(E_1,E_2) = <{\cal D}(E_1){\cal D}(E_2)>
\label{24a}
\end{equation}
which is related to the cumulant as follows

\begin{eqnarray}
\tilde{\cal Y}_2(E_1E_2) && =
- \tilde{\cal R}_2(E_1,E_2) + {\cal R}_1(E_1){\cal R}_1(E_2)\nonumber \\
&& = - <{\cal D}^1(E_1){\cal D}^1(E_2)>
\label{24b}
\end{eqnarray}
\end{mathletters}
For orientation, we mention some properties of this correlator.  We
expect that for large separation of the energies the correlations vanish.
Hence,
\begin{equation}
\int dE\;\; \tilde{\cal Y}_2(E,E{'}) = 0
\label{25}
\end{equation}
In general, we expect only a weak dependence on the absolute values of
the energies; hence, for energies near the Fermi level, the correlator
may be assumed to depend only on the energy difference
\begin{eqnarray}
\tilde{\cal Y}_2(E,E{'}) &&= \tilde Y_2(E-E{'})\nonumber \\
&& =\tilde Y_2(\mid E - E{'}\mid)
\label{27}
\end{eqnarray}
 where the second line follows from the symmetry $E\leftrightarrow
E{'}$.

Making use of Eq. (25) and then of Eq. (\ref{27}), we arrive
at
\begin{eqnarray}
< (\delta N)^2 > && =  s^2\int dE
dE{'}f(E-\mu^0)f(-E{'}+\mu^0)\tilde Y_2(E-E{'})\nonumber \\
&& = s^22T\int du {1\over e^{u/T+1}}\Big(\ell n {1+e^{u/2T}\over 1 +
e^{-u/2T}}\Big)\tilde Y_2(u)
\label{28}
\end{eqnarray}
In the limit $T \rightarrow 0$, the expression simplifies to
\begin{eqnarray}
< (\delta N)^2> = - s^2 \int^0 du u\tilde Y_2(u)\nonumber
\\
= s^2 \int_0 du u\tilde Y_2(u)
\label{29}
\end{eqnarray}
In retrospect, one comment is in order.  We should recognize that the
two energy integrations in the expression for $<(\delta N)^2>$ collect
contributions from a large range of order Fermi energy.  Therefore, it
is not obvious that the translational invariant form (26), for the
correlator, is admissible.  For justification, note that firstly, the
application of the sum rule (25) allows to introduce substantial
restrictions in one energy integration.  Still, the one energy
integration in Eq. (28) may diverge.  There, we should keep in mind that,
we are interested here in quantum effect which  occur at low
energies.
Therefore, it is possible to subtract irrelevant parts and to arrive
at a convergent expression one is interested in.

For a demonstration, see the discussion on persistent currents, where
a metallic ring is threaded by a magnetic flux $\phi$ (see appendix
A). There, it is
found that the phase sensitive part of $<\Delta F>$ is periodic in
$\phi$ with period ${1\over 2}\phi_0$ where $\phi_0 = 2\pi \hbar c/e$
is the flux quantum.  It is also found that the Fourier components of
$<\Delta F>$, that is, the phase sensitive parts of $<\Delta F>$, are
of the order of the mean level spacing $\Delta$.

\section{Canonical Ensemble by Coulomb Blockade}
\label{sec:can}

Here, we discuss the consequences of the fact that  the electrons
carry the charge $(-e)$ and  that the capacitance $C$ of a
mesoscopic sample is small.  (In case of a spherical geometry, $C$ is
proportional to the diameter.)

Suppose that there is an excess charge $(-e)(N-N_i)$ where $N$ is the
number of the electrons and $N_i$ the effective number of the positive
ions.  Then, the electric potential $\phi$ is non-zero and equal to
\begin{equation}
\phi = C(-e)(N-N_i)
\label{30}
\end{equation}
Accordingly, the single particle levels undergo a change
\begin{mathletters}
\label{eq:31}
\begin{equation}
\epsilon_\lambda \rightarrow \epsilon_\lambda - e\phi
\label{31a}
\end{equation}
which can alternatively be expressed as a change in the chemical
potential
\begin{equation}
\mu \rightarrow \mu + e\phi
\label{31b}
\end{equation}
\end{mathletters}

For a discussion of the thermodynamics of the system, it is most
convenient to use $\phi$ instead of $N$ as an independent variable.
In this case, the charging energy $C\phi^2/2$ appears with a minus
sign in the expression for the thermodynamic potential; thus
\begin{equation}
\Omega(T,\mu,\phi) = \Omega(T,\mu + e\phi) - {1\over 2}C\phi^2
\label{32}
\end{equation}
where the ``free'' thermodynamic potential $\Omega(T,\mu)$ is given by
Eq. (3).

In thermal equilibrium, $\Omega(T,\mu,\phi)$ is maximal with respect to
$\phi$; hence
\begin{equation}
\Omega_C(T,\mu) = {\rm min}_\phi\Omega(T,\mu,\phi)
\label{33}
\end{equation}
and the ensuing argumentation  is similar to the one below Eq. (11).
Clearly, a necessary condition for a minimum is
$\partial\Omega(T,\mu,\phi)/\partial\phi = 0$;  that is
\begin{equation}
-e\Big(N(T,\mu,\phi) - N_i\Big) - C\phi = 0
\label{34}
\end{equation}
where the contribution of the positive ions has been subtracted ``by
hand''.

Following the decomposition (7) of the density of states, we have
$\Omega^0 + \Omega^1$ as well as $N = N^0 + N^1$.  Let us now assume
that
\begin{equation}
N^0(T,\mu,\phi = 0) = N_i
\label{35}
\end{equation}
Note that this assumption corresponds to what has been expressed by
Eq. (15a) where the ``typical'' Fermi level $\mu^0$ is defined;
for convenience, we have omitted the superscript zero and we have
put $\mu = \mu^0$ in (34).

As previously, we consider the fluctuating contribution $N^1$ (which
is proportional to
${\cal D}^1$) as well as $\phi$ to be relatively small.  As a
consequence, Eq. (31) can be expanded as follows
\begin{equation}
\Omega(T,\mu,\phi) =\Omega^0 - N^1\cdot(e\phi) - {1\over 2}
{\partial N^0\over \partial\mu}(e\phi)^2 - {1\over 2}C\phi^2
\label{36}
\end{equation}
where the arguments of the thermodynamic quantities on the right side
are $(T,\mu)$.  Calculating the minimum of (35), we find
\begin{mathletters}
\label{eq:37}
\begin{equation}
\Omega_C = \Omega^0 + \Delta \Omega_C
\label{37a}
\end{equation}
where [8]
\begin{equation}
\Delta\Omega_C(T,\mu) = {1\over 2}{1\over C/e^2 + s{\cal D}_F}
\Big(N^1(T,\mu)\Big)^2
\label{37b}
\end{equation}
\end{mathletters}
In the context of the Thomas-Fermi theory, one may consider the
replacement $C/e^2\rightarrow C/e^2 + s{\cal D}_F$ as a manifestation
of screening.

In the limit where the charging energy $E_C$ is much larger than the
mean level distance, that is
\begin{equation}
E_C = {e^2\over 2C} >> {1\over s{\cal D}_F} = \Delta
\label{38}
\end{equation}
we arrive at
\begin{equation}
\Delta\Omega_C(T,\mu) = {1\over 2}\Delta \Big(N^1(T,\mu)\Big)^2
\label{39}
\end{equation}
which agrees exactly with $\Delta F$ of Eq. (23) if the identity $N^1
= \delta N$ -- see Eq. (22) -- is taken into account.

The physics which has lead us to Eq. (38) has been called in
Ref. [7] global charge neutrality.  We wish to add that it is quite
legitimate -- if not obvious -- to consider the mesoscopic sample connected to
an electron
reservoir.  Of course, differences in the work function between
reservoir and sample -- which occur in a real situation -- may require a proper
redefinition of the typical
Fermi level $\mu^0$.

\section{Pinning of the Fermi Energy to a Single Particle Level}
\label{sec:pin}

The theory which we will put down below is based on a statistical
concept of the filling factors of a Fermi system in the presence of
disorder.
At $T=0$, that is, in the ground state, all single particle levels are
fully occupied which satisfy $\epsilon_\lambda \leq \epsilon_F$; the
remaining levels $(\epsilon_\lambda > \epsilon_F)$ are unoccupied.
Of course, the Fermi energy $\epsilon_F$ has to be chosen such that
the
particle number $N$ which follows from this choice,
satisfies the specifications.

At that point one may ask: what are the specifications?  In a
mesoscopic sample of size, say 10$^3$ atomic distances, is there the
number of electrons exactly 10$^9$, or perhaps 10$^9$ + 10$^2$
(10$^0$, 10$^1$, 10$^3$, ...)? Physical considerations suggest that
the
result should not depend on such details.

Therefore, we propose to average with respect to the particle number
in a reasonable window.  A convenient way to realize such an ensemble
will be to select Fermi-energies at random such that
\begin{equation}
\epsilon_F = \epsilon_{\bar\lambda} + \delta
\label{30}
\end{equation}
 where $\epsilon_{\bar\lambda}$ may be any level in a reasonable
range (which will be specified below);
and where $\delta$ is a positive infinitesimal [9].  The above choice
implies that the levels $\epsilon_\lambda \leq \epsilon_{\bar\lambda}$
only are occupied; and that the remaining ones are empty.
We will call this construction the Fermi-level pinning ensemble
(FLPE).

At that point, one may ask: what is the difference then of this
ensemble and
the grand canonical ensemble?  The answer is implicitly given by
arguments found in Ref. [1].  Accordingly, the grand canonical
ensemble (GCE at $T=0$) may be characterized by a random selection of
the chemical potential.  This implies that the transition from the
last occupied level to the first unoccupied one will preferably take
place when the energetic distance is large.  In contrast to it, the
ensemble (FLPE) introduced above
guarantees that such a transition  occurs between any
pair of levels with equal probability.

Again, we start from the expression (1) for the grand canonical
ensemble;
however, we wish to add that only the limit $T\rightarrow 0$ (ground
state) is, strictly speaking, consistent with the argumentation above.
Inserting relation (39), we obtain the following relation for the
``pinned'' grand canonical potential
\begin{equation}
\Omega_P = \sum_\lambda (-T)\ell n \Big[1 + e^{-(\epsilon_\lambda -
\epsilon_{\bar\lambda}-\delta)/T}\Big]
\label{31}
\end{equation}

As a rule, the average properties of a Fermi-system do not depend on
the absolute values of the single particle energies. Therefore, we
remove the arbitrariness in the selection of the pinning level
$\epsilon_{\bar\lambda}$ by sampling with a weight function
$P(\epsilon_{\bar\lambda})$ which is centered at the typical Fermi
energy, that is, $\epsilon_{\bar\lambda} = \mu^0$ (see Eq. (15))
and which has a support much smaller than $\mu^0$ but much larger than
the mean level spacing $\Delta$.  Thus, we obtain for the average
potential
\begin{equation}
<\Omega_P> = {{}\over
<\Sigma_{\bar\lambda}P(\epsilon_{\bar\lambda})>}
<\sum_{\lambda\bar\lambda}P(\epsilon_{\bar\lambda})(-T)\ell n
\Big[1+e^{-(\epsilon_\lambda - \epsilon_{\bar\lambda}-\delta)/T}\Big]>
\label{32}
\end{equation}
Note that we have taken the average separately for numerator and
denominator.
One may justify this procedure in view of the assumed properties of
the weight function $P(\epsilon_{\bar\lambda})$. Nevertheless, this
procedure may give rise to delicate questions since the
properties we are interested in may depend on one electron only.

Inserting the definition (2) of the density of states, and observing
the relations (7) and (24), we obtain

\begin{eqnarray}
<\Omega_P> && = {s\over \int dE {\cal R}_1(\bar E)P(\bar
E)}
\int d E d\bar E\;P(\bar E)\;\tilde {\cal R}_2(E,\bar E)
(-T)\ell n \Big[1 + e^{-(E-\bar
E-\delta)/T}\Big]\nonumber \\
&& = \Omega^0 + <\Delta\Omega_P>
\label{33}
\end{eqnarray}
where $\Omega^0$ is defined by Eq. (8) and where
\begin{mathletters}
\label{eq:43}
\begin{equation}
<\Delta\Omega_P> = - {s\over \int d\bar E{\cal R}_1(\bar E)P(\bar E)}
\int d E d \bar E P(\bar E)\tilde{{\cal Y}}_2(E,\bar E)(-T)\ell n
\Big[1+e^{-(E-\bar E-\delta)/T}\Big]
\label{34}
\end{equation}
By and large, the $E$ integration above collects contributions from a
large range of order Fermi energy.  Therefore, we may not insert at
once the translational invariant form (26) for the correlator.  On the
other hand, the $\bar E$ integration covers only the small range
$\mid \bar E - \mu^0\mid \sim \Delta$ and there, we may assume an
appropriate form of translational invariance.  Thus, we may write
\begin{eqnarray}
<\Delta\Omega_P> && = {-s\over {\cal D}_F\int d\bar EP(\bar E)}\int dEd\bar E
P(\bar E)\tilde{\cal Y}_2(E,\mu^0 - \delta)(-T)\ell n\Big[1 +
e^{-(E-\mu^0)/T}\Big]\nonumber \\
&& = {-s\over {\cal D}_F}\int dE \tilde{\cal Y}_2(E,\mu^0 + \delta)(-T)
\ell n\Big[1+e^{-E-\mu^0)/T}\Big]
\label{34a}
\end{eqnarray}
\end{mathletters}
Note that for $T\rightarrow 0$, we have $(-T)\ell n
[1+e^{-(E-\mu^0)/T}]
\rightarrow \theta(\mu^0 - E)(\mu^0-E)$.  Therefore, the small
quantity $\delta$ is irrelevant.  Furthermore, since we are interested
only in quantum effects, which occur at low energies (``phase sensitive
contributions''), we may now make use of the translation invariant form (26).
Introducing the integration variable
$u = E - \mu^0$ we arrive at
\begin{equation}
<\Delta\Omega_P> = -\Delta s^2\int du \tilde Y_2(u)(-T)\ell n
\Big[1+e^{-u/T}\Big]
\label{35}
\end{equation}
Surprisingly, we recognize that
\begin{equation}
<\Delta\Omega_P (T=0)> = -\Delta s^2 \int^0 du u \tilde Y_2(u)
\label{36}
\end{equation}
is just twice as large as $<\Delta F(T = 0)>$ as given by
Eqs. (\ref{24})
and (\ref{29}).
This apparent discrepancy can be
understood as follows.
We recall that essentially one is interested in the dependence of
$\Omega = \Omega (y)$ on an external parameter $y$.  Such a dependence
is given when the single particle energies $\epsilon_\lambda =
\epsilon_\lambda (y)$ also depend on this parameter.  Thus
\begin{equation}
{\partial\Omega\over \partial y} = \sum_\lambda {\partial
\epsilon_\lambda\over \partial y} f (\epsilon_\lambda - \mu) =: X(y)
\label{37}
\end{equation}

Clearly, the y-dependence will also show up in the correlation
functions.
Therefore, we define
\begin{equation}
\tilde{\cal Y}_2(\bar E,\bar y;E,y) = -<{\cal D}^1(\bar E,\bar y){\cal
D}^1(E,y)>
\label{38a}
\end{equation}
where
\begin{equation}
{\cal D}(E, y) = {1\over s}\sum_\lambda \delta (E-\epsilon_\lambda(y))
\label{38b}
\end{equation}
and ${\cal D}^1 = {\cal D} - {\cal D}^0$.  (We repeat that
${\cal D}^0 = <{\cal D}>$ is not expected to depend on $y$ in an essential
way.) For energies not far from the Fermi energy, the correlator
will be stationary with respect to energy
\begin{eqnarray}
\tilde {\cal Y}_2 (\bar E,\bar y;E,y) && = \tilde Y_2(\bar E-E;
\bar y,y)\nonumber \\
&& = \tilde Y_2(E - \bar E; y,\bar y)
\label{38c}
\end{eqnarray}
and thus, we may generalize Eq. (\ref{35}) as follows
\begin{equation}
<\Delta\Omega_P(y,\bar y)> = -\Delta s^2 \int du \tilde Y_2(u;y,\bar y)(-T)\ell
n \Big[1+e^{-u/T}\Big]
\label{39}
\end{equation}
According to what has been said above, the quantity we are interested
in is
\begin{equation}
< X(y) > = {\partial\over \partial y}<\Omega_P
(y,\bar y)>\Biggr|_{_{\displaystyle{\bar y = y}}}
\label{39a}
\end{equation}
Now, if $\Omega_P$ were symmetric in $y,y{'}$, we would be able to write
\begin{equation}
< X(y) > = {1\over 2} {\partial\over \partial y} <\Omega_P (y,y)>
\label{39b}
\end{equation}
and to remove thus the discrepancy, at least as far as physically
measurable quantities were concerned.

Comparing the structure of $<\Delta\Omega_P>$ and of $<\delta
N(\bar y) \delta N(y)>$, one recognizes that such a symmetry
exists at $T=0$.  For $T\not= 0$, one finds such a symmetry for
$\tilde Y_2(E-\bar E;y,\bar y)$ in the case of a metal
ring threaded by a flux (persistent currents;
see appendix A).  Of course, one should always keep in mind that both
procedures (Legendre transformation and pinning of the Fermi-level)
have a sound basis only for $T=0$.

In conclusion, we have to point out one peculiarity of
the Fermi-level pinning
ensemble. We should not expect any change in  the final
result  if the selection $\epsilon_F =
\epsilon_{\bar\lambda}(y) + \delta$ of Eq. (39) is replaced by
$\epsilon_F = \epsilon_{\bar\lambda}(\bar y) + \delta$.
This expectation is supported by the fact that the energy levels as a function
of $y$ do not cross in disordered (``chaotic'') systems.  Therefore, the set of
levels defined by $P(\epsilon_{\bar \lambda}(y))\gtrsim 0$ should be
the same as $P(\epsilon_{\bar\lambda}(\bar y))$ provided that the weight
function $P$ is sufficiently broad on a range
$\mid\epsilon_{\bar\lambda}(y) - \epsilon_{\bar\lambda}(\bar y)\mid\sim\Delta$.
However, one finds in the case of persistent currents such chances
(see appendix A). Presently, we cannot understand this dependence on such
a detail in the preparation of the Fermi-level pinning ensemble.

For sake of completeness, we wish to mention in this context that
Kamenev and Gefen [10] have discussed a gedanken experiment, where the
system is prepared in a grand canonical state at $\bar y$ whence it is
transferred adiabatically to the parameter value $y$ (see the
discussion in appendix A).

\section{Photoabsorption with Fermi Level Pinned}
\label{sec:photo}

In what follows, we will discuss the consequences of the ansatz (39)
when inserted into relation (\ref{6}) for the photoabsorption.
In a first step,  we
replace $\mu = \epsilon_F = \epsilon_{\bar\lambda} + \delta$ in the
argument
of the Fermi functions.  Next, we take  an average with respect
to the pinning levels by sampling with the weight function
$P(\epsilon_{\bar\lambda})$.
Thus, we obtain the following expression for the photoabsorption
\begin{eqnarray}
\sigma && = {\sigma_0\hbar^2\omega s\over \int d\bar E{\cal D}(\bar
E)P(\bar E)}\int dE dE{'}d\bar E {\cal D}(E){\cal D}(E{'}){\cal
D}(\bar E)P(\bar E)\nonumber \\
&& \times \Big[f (E-\bar E-\delta) - f(E{'} - \bar
E-\delta)\Big]\delta (\hbar\omega - E{'} + E)
\label{40}
\end{eqnarray}

Considering the disorder average $<\sigma >$, we argue as previously, that
it is possible to average numerator and denominator separately.
In this process of
averaging, the three-level correlator appears
\begin{equation}
\tilde{\cal R}_3(E_1,E_2,E_3) = <{\cal D}(E_1){\cal
D}(E_2){\cal D}(E_3)>
\label{41}
\end{equation}
appears.  Since we are interested only in energies close to the typical
Fermi energy $\mu^0$, the correlator ${\cal R}_3$ may be considered to
be invariant with respect to a translation along the energy
axis. Thus, we obtain
\begin{eqnarray}
<\sigma >  && = \sigma_0 {\hbar^2\omega\over {\cal D}_F} s \int
dEdE{'}\tilde{\cal R}_3 (E,E{'},\mu -\delta)\nonumber \\
&& \times\Big[ f(E-\mu^0) - f(E{'}-\mu^0)\Big]\delta
(\hbar\omega - E{'} + E)
\label{42}
\end{eqnarray}

For sake of simplicity, we take the limit $T \rightarrow 0$ where the
Fermi function becomes a step function.  In this case
\begin{equation}
<\sigma > = \sigma_0 {\hbar^2\omega\over {\cal D}_F}
s \int_{-\hbar\omega}^0 du \tilde{\cal R}_3 (u + \mu^0, u +
\hbar\omega
 + \mu^0,\mu^0 - \delta)
\label{43}
\end{equation}
If we were to disregard correlations, ${\cal R}_3 \rightarrow {\cal
D}_F^3$, we would recover Eq. (10).

It will now be necessary to study some general properties of this
correlation function.  Firstly, we define the three level cumulant as
follows
\begin{eqnarray}
\tilde{\cal Y}_3(E_1,E_2,E_3) && = \tilde{\cal
R}_3(E_1,E_2,E_3)\nonumber \\
&& + {\cal R}_1(E_1)\tilde{\cal Y}_2(E_2,E_3)\nonumber \\
&& + {\cal R}_1(E_2)\tilde{\cal Y}_2(E_1,E_3)\nonumber \\
&& + {\cal R}_1(E_3)\tilde{\cal Y}_2(E_1,E_2)\nonumber \\
&& - {\cal R}_1(E_1){\cal R}_1(E_2){\cal R}_1(E_3)
\label{44}
\end{eqnarray}
Again, we expect that for large separation of the energies the
correlations vanish; hence
\begin{equation}
\int dE_3 \;\;\;\tilde{\cal Y}_3(E_1,E_2,E_3) = 0
\label{45}
\end{equation}

It is possible to find definite forms for the correlation function
within the random  matrix theory [11,12].  In our notation, we follow
most closely the article of Bohigas [12].

However, there is one feature, the autocorrelation namely, which does
not seem to have received sufficient
  attention in the past.  For instance, consider Eq. (\ref{24a}) and
insert definition (\ref{2}).  One obtains
\begin{eqnarray}
\tilde{\cal R}_2(E,E{'}) && = {\cal R}_2(E,E{'}) + \delta
(E-E{'}){\cal R}_1(E)\nonumber \\
{\cal R}_2(E,E{'}) && = {1\over s^2}\sum_{\lambda\not=
\lambda{'}}
<\delta (E-E_\lambda)\delta (E-E_{\lambda{'}})>
\label{45a}
\end{eqnarray}
Correspondly, Eq. (24b) is of the form
\begin{eqnarray}
\tilde{\cal Y}_2(E,E{'}) && = - \tilde{\cal R}_2(E,E{'}) + {\cal R}_1
(E){\cal R}_1(E{'})\nonumber \\
&& = {\cal Y}_2(E,E{'}) - \delta(E-E{'}){\cal R}_1(E)
\label{45b}
\end{eqnarray}
such that
\begin{equation}
\int dE{'} {\cal Y}_2(E,E{'}) = {\cal R}_1(E)
\label{45c}
\end{equation}
Autocorrelations play an important role in photoabsorption.  In
thermodynamics, however, they are irrelevant as one may convince
oneself by inspection of Eq. (28) and Eq. (44).

For energies close to the
center of the energy band, that is, for energies close to the Fermi
level
\begin{equation}
 \begin{array}{cc}
{\cal E} & = E-\mu^0\\
{\cal E}{'} & = E{'} - \mu^0
\end{array}
\bigg   \}
\simeq 0
\label{46a}
\end{equation}
we have
\begin{equation}
{\cal R}_1(E\simeq \mu^0) = R_1 ({\cal E} \simeq 0) = {\cal D}_F
\label{46b}
\end{equation}
Furthermore
\begin{eqnarray}
\tilde{\cal Y}_2(E,E{'}) && = \tilde Y_2({\cal E} - {\cal
E}{'})\nonumber\\
\tilde Y_2 ({\cal E}-{\cal E}{'}) && = Y_2({\cal E} - {\cal E}{'}) -
\delta
({\cal E} - {\cal E}{'})R_1(0)
\label{47a}
\end{eqnarray}
where the following rules of notation have been introduced:
Correlators with script letters: arbitrary energies $E_i$;
correlators with roman
letters: small energies ${\cal E}_{i}$.  In this limit, the
correlators are  invariant with respect to
translations. Correlators with (without) tilde: with (without)
autocorrelations.

For reasons of simplicity, the following realtions pertain to the
Gaussian unitary ensemble (GUE) of the random matrix theory. There,
one has
\begin{mathletters}
\label{eq:64}
\begin{equation}
Y_2({\cal E} - {\cal E}{'}) = R_1^2(0)s^2(x-x{'})
\label{64a}
\end{equation}
where
\begin{eqnarray}
s(x) && = {\sin \pi x\over \pi x}\nonumber \\
x_i && = {{\cal E}_i\over s\Delta}
\label{64b}
\end{eqnarray}
\end{mathletters}
We convince ourselves by explicit calculation that
\begin{equation}
\int d{\cal E}{'}\;\;Y_2({\cal E} - {\cal E}{'}) = R_1(0)
\label{48}
\end{equation}
as it should be.

A detailed inspection shows that the three level cumulant of
Eq. (57)
\begin{mathletters}
\label{eq:66}
\begin{equation}
\tilde{\cal Y}_3(E_1,E_2,E_3) = \tilde Y_3({\cal E}_1, {\cal E}_2,
{\cal E}_3)
\label{66a}
\end{equation}
comprises the following $\delta$-function contributions
\begin{eqnarray}
\tilde Y_3({\cal E}_1, {\cal E_2}, {\cal E}_3) = && Y_3 ({\cal E}_1,
{\cal E}_2, {\cal E}_3)\nonumber \\
&& - \delta ({\cal E}_1 - {\cal E}_2)\tilde Y_2({\cal E}_2 - {\cal
E}_3)\nonumber \\
&& - \delta ({\cal E}_2 - {\cal E}_3)\tilde Y_2({\cal E}_3 - {\cal
E}_1)\nonumber \\
&& - \delta ({\cal E}_3 - {\cal E}_1)\tilde Y_2 ({\cal E}_1 - {\cal
E}_2)\nonumber \\
&& - 2\delta({\cal E}_1 - {\cal E}_2)\delta ({\cal E}_2 - {\cal E}_3)R_1(0)
\label{66b}
\end{eqnarray}
\end{mathletters}

Using the above decomposition, we conclude that Eq. (58) assumes the form
\begin{equation}
\int d {\cal E}_3\;\; Y_3 ({\cal E}_1, {\cal E}_2, {\cal E}_3) = 2Y_2
({\cal E}_1 - {\cal E}_2)
\label{48c}
\end{equation}
For the GUE specifically, we have
\begin{equation}
Y_3({\cal E}_1, {\cal E}_2, {\cal E}_3) = R_1^3(0) 2s(x_1 - x_2)s (x_2
- x_3)s(x_3 - x_1)
\label{49a}
\end{equation}
For a control, we calculate
\begin{equation}
\int dx_3 s(x_2-x_3)s(x_3 - x_1) = s (x_1 - x_2)
\label{49b}
\end{equation}
and thus, we find Eq. (68) confirmed.

In accordance with the notation introduced above,  we put
\begin{mathletters}
\label{eq:70}
\begin{equation}
\tilde{\cal R}_3 (E_1, E_2,E_3) = \tilde R_3({\cal E}_1, {\cal E}_2,
{\cal E}_3)
\label{70a}
\end{equation}
and find the explicit form
\begin{eqnarray}
\tilde R_3({\cal E}_1, {\cal E}_2, {\cal E}_3) && = Y_3 ({\cal E}_1,
{\cal E}_2, {\cal E}_3)\nonumber \\
&& - R_1(0)\Big\{Y_2({\cal E}_2 - {\cal E}_3) + Y_2 ({\cal E}_1 -
{\cal E}_3) + Y_2 ({\cal E}_1 - {\cal E}_2)\Big\}\nonumber \\
&& + \delta ({\cal E}_1 - {\cal E}_2) \Big[R_1^2(0) - Y_2 ({\cal E}_2
- {\cal E}_3)\Big]\nonumber \\
&& + \delta ({\cal E}_2 - {\cal E}_3)\Big[R_1^2(0) - Y_2 ({\cal E}_3 -
{\cal E}_1)\Big]\nonumber \\
&& + \delta ({\cal E}_3 - {\cal E}_1)\Big[R_1^2(0) - Y_2 ({\cal E}_1 -
{\cal E}_2)\Big]\nonumber \\
&& + R_1(0)\delta ({\cal E}_1 - {\cal E}_2)\delta ({\cal E}_2 - {\cal
E}_3)+ R_1^3(0)
\label{70b}
\end{eqnarray}
\end{mathletters}

We insert this form in the relation (56) for photoabsorption.
Since
$<\sigma>$ is an even function of $\omega$, we take $\omega > 0$.
Firstly, we recognize that the contribution with a product of two
$\delta$-function in Eq. (71b) drops out.  Next, we also
recognize that the infinitesimal $\delta$ is important for an exact
definition of the contributions, which  the remaining $\delta$-functions
collect from  the end point of the integration interval.

Using the specific forms of Eq. (65) and Eq. (69) and
normalizing the cross-section $<\sigma>$ to the uncorrelated one of
Eq. (10), we may put the result in the following form
\begin{mathletters}
\label{eq:71}
\begin{eqnarray}
{<\sigma >\over <\sigma >^{uc}} && = {2\over x} s(x) \int_{-x/2}^{x/2}
dy\;\; s(y + {x\over 2})s(y - {x\over 2})\nonumber \\
&& - s^2(x) + {2\over x} - {2\over x}s^2(x) - {1\over x} \int_{-x}^x
dy\;\;s^2(y) + 1
\label{71a}
\end{eqnarray}
where
\begin{equation}
x = {\hbar \omega\over s\Delta}
\label{71b}
\end{equation}
\end{mathletters}
For small and large frequencies, the above relation reduces to
\begin{equation}
{<\sigma >\over <\sigma >^{uc}} = \cases{(2\pi^2/3)\hbar\omega/s\Delta
 & $\hbar\omega << \Delta$\cr
1 + s\Delta/\hbar\omega & $\hbar \omega >> \Delta$\cr}
\label{51c}
\end{equation}
By numerical integration, we have calculated the cross-section for
intermediate frequencies.  The result is shown in Fig. 1

The same type of analysis can also be done for the
Gaussian Orthogonal Ensemble (GOE) of the random matrix theory.
This is outlined in appendix B.  In this case
\begin{equation}
{<\sigma >\over <\sigma >^{uc}} =
\cases{(\pi^2/3)-(\pi^4/30)(\hbar\omega/s\Delta)^2 & $\hbar
\omega << \Delta$\cr 1 + s\Delta/\hbar\omega & $\hbar\omega >> \Delta$\cr}
\label{73}
\end{equation}
For intermediate frequencies, the cross-section is found by numerical
integration; for a graphical representation see  Fig. 1.

The limiting forms (73) and (74) are in agreement with Shklovskii's
[1] conclusions.

\section{Stochastic Fluctuations}
\label{sec:stoch}

Up till now, it seems to be an accepted procedure in mesoscopic
physics to calculate stochastic fluctuations within the grand
canonical ensemble.  Thus, one calculates the connected correlator of
the grand canonical potentials (see also Eq. (47))
\begin{eqnarray}
<\Omega(T,\mu,y)&&
\Omega (T,\mu,y{'})>_c  =
<\Omega\Omega{'}>-<\Omega><\Omega{'}>\nonumber \\
&& = -s^2\int dEdE{'} \tilde{\cal Y}_2(E,y;E',y')T^2\Big(\ell n\Big[1
+
e^{-(E-\mu)/T}\Big]\Big)\Big(\ell n\Big[1 +
e^{-(E{'}-\mu)/T}\Big]\Big)
\label{75}
\end{eqnarray}
According to Eq. (46), the correlator $<X(y)X(y{'})>_c$ for the
physical quantity $X(y)$ can be obtained by operating with
$\partial^2/\partial y\partial y{'}$ on Eq. (75).

As an illustration, such a calculation is presented in appendix A for
the case of persistent currents.  One obtains for the phase sensitive
part of $<\Omega\Omega{'}>_c$ by order of magnitude $\sim E_c$, where
$E_c = D/L^2$ is the Thouless energy in the diffusive limit.  Note that
$E_c >> \Delta$ in experiments of the usual type.

However, with the results and with the insight obtained in the
discussions of the preceding sections, we may quite well ask what the
stochastic fluctuations are when calculated for the Fermi level
pinning ensemble.  Considering Eq. (40), we write down in a first step
\begin{equation}
\Omega_P(y,\bar y)\Omega_P(y{'},\bar y) = \sum_{\lambda\lambda{'}}T^2
\Big(\ell n \Big[1 + e^{-(\epsilon_{\bar\lambda}(\bar
y)+\delta)/T}\Big)
\Big(\ell n\Big[1 +
e^{-(\epsilon_{\lambda{'}}(y{'})-
\epsilon_{\bar\lambda}(\bar y)+\delta)/T}\Big]\Big)
\label{76}
\end{equation}
Next, we perform an average of $\epsilon_{\bar\lambda}$ in a
reasonable energy range on the basis of the sampling function
$P(\epsilon_{\bar\lambda})$.  As previously, we claim that it is
possible to take the disorder average for numerator and denominator
separately.  Thus, we obtain
\begin{eqnarray}
<\Omega_P(y,\bar y) && \Omega_P(y{'},\bar y)>\nonumber \\
&& = {s^2\over {\cal D}_F\int d\bar EP(\bar E)}\int dEdE{'}d\bar EP(\bar
E)<{\cal D}(E,y){\cal D}(E{'},y{'}){\cal D}(\bar E,\bar y)>\nonumber \\
&& \times T^2\Big(\ell n\Big[1 + e^{-(E-\bar E)/T}\Big]\Big)\Big(\ell
n\Big[1 +e^{-(E{'}-\bar E)/T}\Big]\Big)
\label{77}
\end{eqnarray}

As far as the energy integration is concerned, we wish to recall the
argumentation in connection with Eq. (43). Accordingly, we
write (typical Fermi-energy $\mu^0 \rightarrow \mu$)
\begin{eqnarray}
<\Omega_P(y,\bar y)\Omega_P(y{'},\bar y)> && = {s^2\over {\cal D}_F}\int
dEdE{'}<{\cal D}(E,y){\cal D}(E{'},y{'}){\cal D}(\mu + \delta,\bar y)>\nonumber
\\
&& \times T^2\Big(\ell n\Big[1 + e^{-(E-\mu)/T}\Big]\Big)\Big(\ell n
\Big[1+e^{-(E{'}-\mu)/T}\Big]\Big)
\label{78}
\end{eqnarray}

Since $<\Omega(\mu,y)\Omega(\mu,y{'})>$ can be obtained rather easily,
it is advantageous to calculate the difference
\begin{eqnarray}
<\Omega_P(y,\bar y)\Omega_P(y{'},\bar y)> && -
<\Omega(\mu,y)\Omega(\mu, y{'})>\nonumber \\
&& = s^2\int dEdE{'} \Big\{{1\over {\cal D}_F}\tilde{\cal
R}_3(E,y;E{'},y{'};\mu + \delta,\bar y)-\tilde{\cal
R}_2(E,y;E{'},y{'})\Big\}\nonumber \\
&& \times T^2\Big(\ell n\Big(\ell n\Big[1 +
e^{-(E-\mu)/T}\Big]\Big)\Big(\ell n \Big[1 + e^{-(E{'}-\mu)/T}\Big)
\label{79}
\end{eqnarray}
For a definition of $\tilde{\cal R}_2$ and $\tilde{\cal R}_3$, see
Eqs. (24) and (54), respectively.  As far as the difference of the
correlators in the curly brackets is concerned, we make use of
Eq. (57) and obtain
\begin{eqnarray}
{1\over {\cal D}_F}\tilde{\cal R}_3(E,y;E{'},y{'};\mu + \delta,\bar y) && -
\tilde{\cal R}_2(E,y;E{'},y{'})\nonumber \\
&& = {1\over {\cal D}_F}\tilde{\cal Y}_3(E,y;E{'},y{'};\mu + \delta,\bar
y)\nonumber \\
&& - {1\over {\cal D}_F}R_1(E)\tilde{\cal Y}_2(E{'},y{'};\mu + \delta,\bar y)
- {1\over {\cal D}_F}(E{'})\tilde Y_2(E,y;\mu + \delta,\bar y)
\label{80}
\end{eqnarray}
As emphasized repeatedly, we are considering systems where $R_1(E) =
<{\cal D}(E,y)>$ does not depend on the external parameter $y$.

We insert the last term of Eq. (80) in Eq. (79) and find
\begin{eqnarray}
&& - {s^2\over {\cal D}_F}\int dEdE{'}{\cal R}_1(E{'})\tilde{\cal Y}_2
(E,y;\mu -
\delta,\bar y)T^2\Big(\ell n\Big[1 + e^{-(E-\mu)/T}\Big]\Big)\Big(\ell
n\Big[1+e^{-(E{'}-\mu)/T}\Big]\Big)\nonumber \\
&& = \Omega^0(\mu)<\Delta \Omega_P(y,\bar y)>
\label{81}
\end{eqnarray}
The last line follows from Eqs. (8), (43) and (50).  Similarly, the
second last term of Eq. (8) contributes with
$\Omega^0(\mu)<\Delta\Omega_P(y{'},\bar y)>$.

At this, we recall that ultimately, we are interested in the
correlator of the physical quantity $X(y)$.  This means that we should
calculate
\begin{equation}
<X(y)X(y{'}(>_{P,\bar y} = {\partial^2\over \partial y\partial
y{'}}<\Omega_P(y,\bar y)\Omega_P(y{'},\bar y)>
\label{82}
\end{equation}
(The subscript $P$ indicates how the average is taken and $\bar y$ is
the parameter value at which the Fermi level pinning ensemble has been
prepared.) It is now important to realize that terms of the type (81)
do not contribute to the correlator (82).

For sake of transparent formulae, we will omit terms of the type (81);
in the following equalities are meant  to be relations of
equivalence in the sense of Eq. (82).
Therefore, we may write
\begin{eqnarray}
&& <\Omega_P(y,\bar y)\Omega_P(y{'},\bar y)>  -
<\Omega(\mu,y)\Omega(\mu,y{'})>\nonumber \\
&& = {s^2\over {\cal D}_F}\int dEdE{'}\tilde{\cal Y}_3\Big(E,y;E{'},y{'};\mu
+ \delta,\bar y\Big)T^2\Big(\ell
n\Big[1+e^{-(E-\mu_/T}\Big]\Big)\Big(\ell n
\Big[1 + e^{-(E{'}-\mu)/T}\Big]\Big)
\label{83}
\end{eqnarray}

Note that for $T\rightarrow 0$, the last factors in the integrand are
${\cal E}{\cal E}{'}\theta(-{\cal E})\theta(-{\cal E}{'})$ where
${\cal E},{\cal E}{'}$ are energies measured from the Fermi level as
defined in Eq. (62).  Clearly, large values of $\epsilon,\epsilon{'}$
are important in the integral (83), and a simple RMT theory fails if
not a prescription how to extract phase sensitive contribution is
supplied. As an alternative, one may resort to the diagrammatic theory
and to the cooperon-diffuson expansion.  A preliminary analysis
suggests that the expression (83) is of the order
$E_c^2(L/\ell)^2(\Delta^2/\gamma E_c)$
where $L$ is the circumference and $\ell$ the mean free path of the
metallic ring. Furthermore, $\gamma$ the rate at which phase coherence
is destroyed.
Since $<\Omega(\mu,y)\Omega(\mu,y{'})> \sim E_c^2$,
the estimate above  indicates that in metals of short mean free path
the stochastic
fluctuations may be  considerably larger  within the
Fermi level pinning ensemble,
that is, when calculated for a fixed number of electrons as compared
with calculations for fluctuating number of electrons.

\section{Discussion}
\label{sec:discuss}

We recall the discussion on the existence and on the size
of persistent currents in mesoscopic metallic rings threaded by a
magnetic flux $\phi$ which took place just a few years ago [2].  It
has been found that for non-interacting electrons, the persistent
currents are exponentially small [13]  when calculated  for
the grand canonical ensemble (GCE)  in contrast to calculations for
the canonical ensemble (CE).  There one finds persistent currents,
 (when expressed in terms of energies)  of
order  $\Delta$, that is, the mean spacing of the electronic
level. Nevertheless, even this result is small and it is presumably
correct to say that the size of the persistent currents depends on
just one electron.

In such a case, it is necessary to have reliable information on the
distribution of the electronic levels $\epsilon_\lambda$.  A basic
assumption is that atomic disorder or fluctuations in the geometric
definitions exist in all samples which are
prepared identically on a mesoscopic scale.  Correlation functions
(with respect to this disorder), of these levels can be calculated in
an approximation by a diagrammatic method [14] or in the random matrix
theory [11,12].  A comprehensive approach is provided by the
supersymmetry technique [15], but technically, it is not so easy to
handle.

Two main subjects are discussed in this paper: (A) Thermodynamics; (B)
Dynamics; (C) Stochastic Fluctuations.

Concerning (A), there is no need to emphasize, that  for
systems of identical particles, a theory can be formulated and carried
through  most  elegantly in the
grand canonical ensemble.  Thus, there is a possibility (i) to extract
from this information thus obtained the canonical ensemble by  a Legendre
transformation.  In a second procedure (ii), one starts from the idea of a
Coulomb blockade where the charging energy $e^2/2C$ ($C$ is the
capacitance of the mesoscopic sample) fixes the number of electrons
to be equal to the (effective) number of ions provided that $e^2/2C >>
\Delta$.  A third possibility (iii) is based on the concept of
pinning the Fermi level to a single particle level.  In this context,
one argues that effectively the grand canonical ensemble can be
understood as  a
random selection of Fermi energies which prefers
configurations where the energy separation between last occupied and
first empty state is large. In contrast to it, the Fermi level
pinning ensemble provides an unbiased choice.

By and large we have found that (at $T=0$) the
three methods do lead to the same result as far as the thermodynamics
is concerned.  One open question remains.  In terms of persistent
current: Is there a difference in the
persistent currents of a flux $\phi$ when the system has been prepared
at a flux $\bar\phi \not= \phi$?  Some comments can be found at the
end of appendix A. In this context, the following problem (which at
first sight seems to be none at all),  should also be considered.  We recall
Eq. (15) where $\mu^0$ is defined by $-\partial \Omega^0/\partial \mu
\mid_{\mu = \mu^0}=N$.  A relation of similar structure is Eq. (34).
The question is: how is it possible to find $\mu^0$ such that $N$ is
an integer?  Clearly, within the present formalism, we cannot
guarantee $N$ to be an integer.  On the other hand, such a detail may
be very important since we have convinced ourselves that it may be only
one electron which contributes to the phenomenon we are interested in.

As an example of dynamics (B), we have studied the disorder average
cross section $<\sigma(\omega)>$ for the photoabsorption of a
mesoscopic sample as proposed in Ref. [1].  According to a qualitative
analysis of Shklovskii [1], the cross section $<\sigma(\omega)>$ is
reduced by a factor $(\hbar\omega/\Delta)$ ($<< 1$ for small
frequencies) as compared with a naive calculation in the grand
canonical ensemble.  We have
calculated $<\sigma(\omega)>$ using the concept (iii) of Fermi level
pinning.  There, we have obtained agreement with Shklovskii's ideas
within the random matrix
theory.  Furthermore, the present theory allows us to cover all frequencies.

It seems to be tempting to calculate photoabsorption within the
Coulomb blockade concept (ii).  However, in a naive treatment,
correlation functions of very high order $(\rightarrow \infty)$ seems
to be required; a situation which calls for a more detailed
discussion.  In this context, we wish to mention also an approach to
this dynamic problem which has been put forward by Kamenev et
al. [16], which agrees with the limiting form proposed by Shklovskii
[1].

Stochastic fluctuations (C) seem to be very important in mesoscopic
physics. Though the average response may be small, there are in
general large sample to sample fluctuations of this response.  As a
simple measure of these fluctuations, the root-mean-square of the
response is most useful.  Usually, the r.m.s is calculated within the
grand canonical ensemble.  We have studied this quantity for Fermi
level pinning ensemble (iii).  In the case of persistent currents, we
have outlined an estimate which indicates a considerable enhancement
is possible in the
fluctuations of persistent currents, when calculated for fixed number
of electrons.

At the end, we wish to draw attention to measurements where the
magnetic moment of (singly connected) mesoscopic samples has been
measured by L\'evy et al. [17].  Again, the question of differences
between canonical and grand canonical ensemble arises.  In the
presence of strong disorder (diffusive limit), the paper by Altshuler
et al. [18], seems to provide a complete answer.  On the other hand, if
there is only weak atomic disorder, the problem is more complicated [19].  A
new physical situation arises when samples are free from intrinsic
defects (Aharonov-Bohm ballistic billiards), and where disorder
appears only in the form of  differences in the  geometric definitions
of the samples.
This situation has been discussed by Ullmo et al. [20].  We, together
with Yu. N. Ovchinnikov, are also working on this problem,
analytically as well as numerically.  It seems to be a very difficult
problem; one could say, that it means to catch the one active electron
among the million inert ones.

\acknowledgments

A.S. wishes to thank A. Kamenev and Y. Gefen for stimulating
discussions during his stay at the Weizmann Institute.  In fact, these
discussions have initiated the investigation of this paper.
In addition, it is a pleasure to acknowledge illuminating comments by
A. Mirlin. This
work has been supported by the German-Israeli Foundation (GIF).

\appendix
\section{}

According to Altshuler and Shklovskii [14], the two level correlation
functions can be expanded in two types of impurity ladder diagrams
that are called diffusons and cooperons.  For this expansion to be
valid,
a condition has to be satisfied which is that the energetic separation
of the two levels to be considered has to be much larger
than the mean level separation
\begin{equation}
\mid E - E{'}\mid >> \Delta
\label{A1}
\end{equation}
In the framework of this diagramatic theory, one obtains a
two-diffuson contribution which is as follows
\begin{equation}
<{\cal D}^1(E){\cal D}^1(E{'})>^{2D} = {1\over 2\pi^2}\sum_{\vec q}
{\rm Re}\;{1\over [-i(E-E{'})+\gamma+D\vec q^2]^2}
\label{A2}
\end{equation}
In the relation above, $D\vec q^2$ represents the eigenvalues of the diffusion
operator
$-D\vec\nabla^2$ for the sample in consideration and  $\gamma$
means the rate at which phase coherence is destroyed.  In case of
time reversal symmetry, the two-cooperon contributions $<{\cal
D}^1(E){\cal D}^1(E{'})>^{2C}$ is the same as expression (\ref{A2}).

We call a sample to be effectively of zero dimension if the zero
eigenvalue
of $-D\vec\nabla^2$ is the only important one.  In this case
\begin{equation}
<{\cal D}^1(E){\cal D}^1(E{'})>^{2D} = - {1\over 2\pi^2}{\rm
Re}{1\over (E-E{'}+i\gamma)^2}
\label{A3}
\end{equation}
According to the definition (24), the expression (\ref{A3}) has to be
considered as the two-diffuson approximation of the two-level
correlator
$-\tilde{\cal Y}_2(E,E{'})$.  We also observe that this approximation
satisfies the sum rule (25).  Furthermore, it agrees with the GUE
result (65) of the random matrix theory for large energy separation
$\mid E-E{'}\mid >> \Delta;\gamma$, if we replace there the fast
oscillatory term $\sin^2\pi x$ by its average value ${1\over 2}$.
For small energies, however, the agreement is bad.

As mentioned above, the two-cooperon contribution is the same for a
system with perfect time reversal symmetry, $\tilde Y_2^{2C} = \tilde
Y_2^{2D}$.  On the other hand, if time reversal symmetry is completely
lost, the two-cooperon contribution is zero, $\tilde Y_2^{2C} = 0$.
In the random matrix theory, the two cases correspond to GOE and GUE,
respectively.

Despite its deficiency, the cooperon-diffuson expansion is useful
since it can easily be generalized to one  and higher dimensional
samples.  In the discussion of persistent currents in metallic rings
threaded by a magnetic flux $\phi$, one considers frequently an
effectively one-dimensional closed loop of length $L$; as a consequence, the
eigenvalues $D\vec q^2$ of the diffusion operator can be found by
replacing
\begin{mathletters}
\label{eq:A5}
\begin{equation}
\vec q \rightarrow {2\pi\over L}(n + \varphi)
\label{A5a}
\end{equation}
where $\varphi = 0$ for diffusons whereas for cooperons
\begin{equation}
\varphi = {\phi\over \phi_0}\;\;;\;\;\phi_0 = {2\pi \hbar c\over e}
\label{A6b}
\end{equation}
\end{mathletters}

We remark that the ansatz (A4) neglects any penetration of the
magnetic
field into the area of the metallic ring.  Therefore, the cooperon
contribution remains oscillating even for very large flux and no
transition to the GUE ensemble of random matrix theory will take place.

Presently, we are interested in the thermodynamics only in so far
as the flux dependence is concerned. Hence, only the two-cooperon
contribution $\tilde Y_2^{2C}$ matters.  Considering Eqs. (23) and
(28), we recognize that the $\delta$-function contribution to $\tilde
Y_2$ is irrelevant.  We will also see later, that neither the
$\eta$-regularisation of the low energy dependence nor the phase breaking rate
$\gamma$ is not essential in the present
problem.  Thus, we may
write
\begin{equation}
Y_2^{2C}(u;\varphi) = {1\over 2\pi^2}\sum_n {\rm Re}\;{\over
[u+iD(2\pi/L)^2(n+\varphi)]^2}
\label{A6}
\end{equation}
Note that $Y_2^{2C}$ is an even  function of $\varphi$; moreover, it
is as periodic in $\varphi$
with period $1$.  Therefore, a Fourier expansion is appropriate and we put
\begin{eqnarray}
Y_2^{2C}(u;\varphi) && = \sum_mC_m(u)e^{2\pi im\varphi}\nonumber \\
C_m(u) && = \int_{-\infty}^{+\infty} d\varphi\;\; e^{-2\pi i m\varphi} {1\over
2\pi^2} {\rm
Re}{1\over [u+iD(2\pi/L)^2\varphi^2]^2}
\label{A7}
\end{eqnarray}
Explicitly, one finds
\begin{mathletters}
\label{eq:A8}
\begin{equation}
C_m(u) = {1\over 8\pi^2}{1\over E_c^2}{\rm Re}\Big[{e^{-i\pi/4}\over
v^{3/2}}-i {\mid m\mid\over v}\Big]e^{i\mid m\mid{1\over 2} e^{i\alpha(v)}}
\label{A8a}
\end{equation}
where
\begin{equation}
E_c = D/L^2
\label{A8b}
\end{equation}
is the Thouless energy and where
\begin{eqnarray}
v && = u/E_c\nonumber \\
\alpha(v) && = \pi/2 - (\pi/4){\rm sgn}v
\label{A8c}
\end{eqnarray}
\end{mathletters}
Next, we consider Eq. (28) and calculate the cooperon
$(\varphi = 2\phi/\phi_0)$ contribution
\begin{eqnarray}
<(\delta N(\phi))^2> && = \sum_mB_me^{2\pi i m2\phi/\phi_0}\nonumber
\\
B_m && = s^2\int_0 du \;\;uC_m(u) = {s^2\over 2\pi^2}{1\over \mid
m\mid}
\label{A9}
\end{eqnarray}

We observe that the divergence of $C_m(u)\propto u^{-3/2}$ is
irrelevant in the integration and also, that the integral
converges for large values of $u$ if $m\not= 0$.  This last point may
be considered as a demonstration of the rule that only low energies
(here: $\mid u \mid \leq E_c$)
contribute to phase sensitive quantities.

For sake of completeness, we give the expression for the phase
sensitive
part of the free energy.  Inserting expression (\ref{A9}) in Eq. (28),
we have
\begin{mathletters}
\label{eq:10}
\begin{equation}
<\Delta F> = \sum_{m=1}^\infty {s^2\over 2\pi^2}{\Delta\over
m}\cos\;\;2\pi m{2\phi\over \phi_0}
\label{A10a}
\end{equation}
Since the persistent current $I(\phi)$ can be calculated according to
\begin{equation}
I(\phi) = - c {\partial\over \partial\phi} <\Delta F>
\label{A10b}
\end{equation}
\end{mathletters}
we obtain a result that can be expressed as follows
\begin{eqnarray}
I(\phi) && =: {\cal J}(\phi) \nonumber\\
{\cal J}(\phi)    && = \sum_{m=1}^\infty {s^2\over 2\pi^2}\cdot
{2e\over\hbar}\cdot {\Delta\over m}\sin 2\pi m{2\phi\over \phi_0}
\label{A11}
\end{eqnarray}

Following the argumentation which lead to  Eq. (47), we also
define a two-level correlator which depends on the two fluxes $\phi$
and $\bar \phi$.  In this case, cooperons as well as diffusons
contribute to the phase sensitive quantities; and we obtain the
appropriate correlators if we substitute $\varphi = (\phi +
\bar\phi)/\phi_0$ for cooperons and $\varphi = (\phi - \bar\phi)/\phi_0$
for diffusons in expression (\ref{A7}).  (Note the symmetry
$\phi\leftrightarrow \bar\phi$.) From these relations, it follows that
\begin{eqnarray}
<\Delta\Omega_P(T=0;\phi,\bar\phi)> && = \Delta <\delta N(\phi)\delta
N(\bar\phi)>\nonumber\\
    && = \sum_{m=1}^\infty {s^2\over \pi^2} {\Delta\over m}\Big[\cos
2\pi m {\phi +\bar\phi\over \phi_0}+\cos 2\pi m\;{\phi - \bar\phi\over
\phi}\Big]
\label{A12}
\end{eqnarray}
Evidently,
\begin{eqnarray}
I_P(\phi,\bar\phi) && = - c {\partial\over \partial\phi}
<\Delta\Omega_P
(\phi,\bar\phi)> \nonumber\\
    && = {\cal J}\Big({1\over 2}(\phi + \bar\phi)\Big) + {\cal J}
\Big({1\over 2}(\phi - \bar\phi)\Big)
\label{A13}
\end{eqnarray}
As expected, the above result agrees with Eq. (\ref{A11}) only if
$\bar\phi = \phi$.

Next, we comment the paper of Kamenev and Gefen [10].
They consider an experiment where the system is prepared in a grand
canonical state at the flux $\bar\phi$ and then transferred adiabatically
to a state with flux $\phi$.  Their result for the persistent current
can be expressed as follows
\begin{eqnarray}
I(\phi,\bar\phi) && = - c{\partial\over \partial\phi}{1\over 2}\Delta
<(\delta N(\phi) - \delta N(\bar\phi))^2>\nonumber\\
    && = {\cal J}(\phi) - {\cal J}\Big({1\over 2}(\phi + \bar\phi)\Big)
- {\cal J}\Big({1\over 2}(\phi - \bar\phi)\Big)
\label{A14}
\end{eqnarray}
In conclusion, we note the interesting relation
\begin{equation}
I_P (\phi, \bar\phi) + I(\phi,\bar\phi) = {\cal J}(\phi)
\label{A15}
\end{equation}
which we do not understand presently in physical terms.

As far as the problem of stochastic fluctuations is concerned, it is not
difficult to calculate the connected part
$<\Omega(\phi)\Omega(\phi{'})>_c$ of the correlator for the grand
canonical potential in the diffuson-cooperon approximation outlined
above (see also the standard results for noninteracting electrons
quoted in Ref. [3]). Accordingly, one finds for the grand canonical
ensemble at  $T=0$, the following result
\begin{equation}
<\Omega(\phi)\Omega(\phi{'})>_c = \sum_{m=1}^\infty {96\over
\pi^2m^5}E_c^2\cos
{2\pi m\phi\over \phi_0}\cos {2\pi m\phi{'}\over \phi_0}
\label{A16}
\end{equation}
Note that the fluctuations $\sim E_c$ are much larger than the mean
values $\sim\Delta$.

\section{}

We outline our calculation of the disorder averaged
the normalized cross-section (\ref{40})
for the GOE. Making use of Eqs. (\ref{70b}) and (\ref{71b}),
 we obtain the relation
\begin{eqnarray}
{<\sigma >\over <\sigma >^{uc}} && = {1\over x}  \int_{-x}^{0}
dy\;\;  Y_3 ({y}, {y+x}, {0})\nonumber \\
&& -  {1\over x} \int_{-x}^x
dy\;\; Y_2({y})+ \left(1+ \frac{2}{x} \right) \left(1- Y_2(x) \right)
\label{b1}
\end{eqnarray}

In the RMT, the $n$ point cummulant function is given by the expression
\begin{equation}
Y_n({\cal E}_1,
{\cal E}_2,..., {\cal E}_n)=\frac{1}{2} \mbox{Tr} \sum_p
(\sigma(x_{12})\sigma(x_{23})\sigma(x_{n1})),
\label{b2}
\end{equation}
where $ x_{ij}=x_i-x_j$  and $\sum_P$ denotes the sum over the
subsets of the symmetric group with respect to the
group of cycle permutation. We note that in the present problems, these
permutations produce analytic forms which are identical.

The matrices $\sigma(x)$ are given by
\begin{equation}
\sigma(x)=\left(
\begin{array}{cc}
s(x) & D(x)\\
J(x) & s(x)
\end{array} \right)=\left(
\begin{array}{cc}
s(x) & \frac{\partial}{\partial x}s(x)\\
\int_0^x s(t)dt-\frac{1}{2} \mbox{sgn}(x) & s(x)
\end{array} \right),
\label{b3}
\end{equation}
where $s(x)$ is found in Eq. (\ref{64b}).
For the two point cummulant  we get
\begin{equation}
Y_x(x)=s^2(x)-D(x)J(x)
\end{equation}

The three point cummulant can be calculated according to
\begin{eqnarray}
Y_3 ({y}, {y+x}, {0})&=&2s(x)s(y+x)s(y)\nonumber\\
	                  &&+s(x)D(y+x)J(y)+s(x)J(y+x)D(y)\nonumber\\
	                  &&+D(x)s(y+x)J(y)+D(x)J(y+x)s(y)\nonumber\\
	                  &&+J(x)D(y+x)s(y)+J(x)s(y+x)D(y)\label{b4}.
\end{eqnarray}
We insert these quantities in equation (\ref{b1}) and after some
manipulations, we obtain  the limiting
cases, Eq. (\ref{73}). Numerical calculation have been done
accordingly.

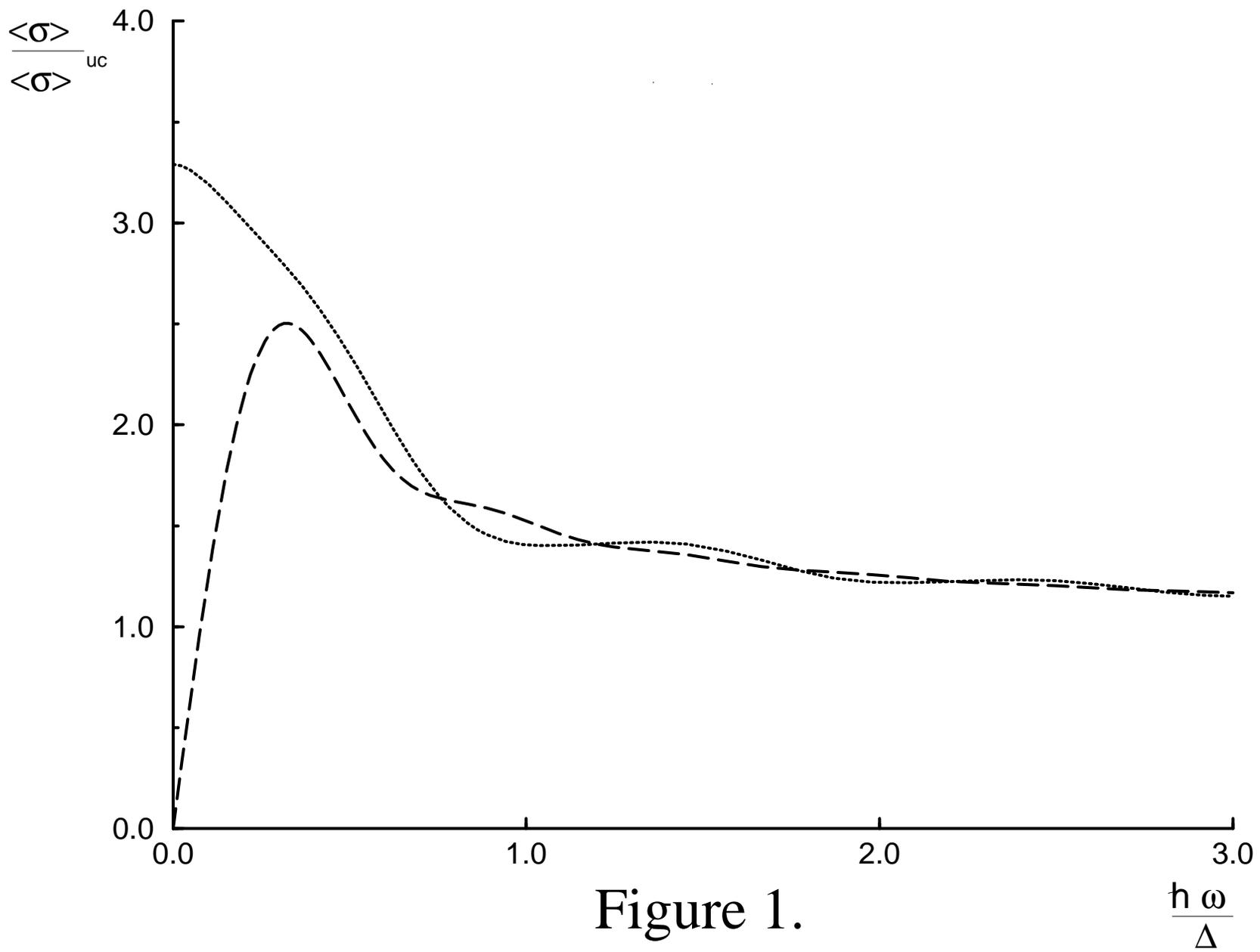
\begin{figure}
\caption{ Average cross section  $<\sigma>$ of photoabsorption for an
ensemble of mesoscopic samples, as a function of the photoenergy
$\hbar \omega$,
 normalized to the mean level distance $\Delta$ (spin deceneracy s=2).
The cross section is given in the form of the ratio
$<\sigma>/<\sigma>^{uc}$,
where $<\sigma>^{uc} \propto \omega^2$ is the uncorrelated quantity.
Dotted line: Gaussian orthogonal ensemble;
dashed line: Gaussian unitary ensemble (GOE and GUE, respectively, of
random matrix theory).}
\label{Fig.1}
\end{figure}
\end{document}